# EAIA: An Efficient and Anonymous Identity Authentication Scheme in 5G-V2V

Qianmin Du[1], Jianhong Zhou[1], Maode Ma[2]
[1]School of Computer and Software Engineering, Xihua University, Sichuan, China
[2] KINDI Computing Research Center, College of Engineering, Qatar University, Qatar
acadmmd@gmail.com

*Abstract*—Vehicle Ad-hoc Networks (VANETs) have experienced significant development in recent years, playing a crucial role in enhancing the driving experience by enabling safer and more efficient inter-vehicle interactions through information exchange. Vehicle-to-vehicle (V2V) communication is particularly vital as it not only helps to prevent collisions and improve traffic efficiency but also provides essential situational awareness to drivers or autonomous driving systems. Communication is typically supported by Roadside Units (RSUs); however, in practical applications, vehicles may exceed the communication range of RSUs, thus exposing them to various malicious attacks. Additionally, considering the limited computational resources of onboard units (OBUs) in vehicles, there is a high demand for designing lightweight security protocols that support V2V communication. To address this issue, this paper proposes an efficient anonymous V2V identity authentication protocol tailored for scenarios that lack RSU support. The proposed protocol has been formally assessed using the Scyther tool, demonstrating its capability to withstand major typical malicious attacks. Performance evaluations indicate that the proposed protocol is efficient in terms of communication and computational overhead, making it a viable solution for V2V vehicle communication.

## I. Introduction

In recent years, the technology of connected vehicles has been evolving towards greater intelligence and network integration[1]. Vehicle-to-Everything (V2X) technology has become a key enabler for the exchange of information in intelligent connected vehicles. V2X technology enhances vehicles' perception of the traffic environment by enabling early access to information such as operational data of nearby vehicles, traffic control information, congestion data, and visual blind spots, thereby facilitating information sharing among vehicles. It encompasses four main types: Vehicle-to-Infrastructure (V2I), Vehicle-to-Vehicle (V2V), Vehicle-to-Pedestrian (V2P), and Vehicle-to-Network (V2N) communication[2]. V2I communication focuses on the interaction between vehicles and road infrastructure to receive local traffic broadcasts. V2V communication primarily involves active safety services related to communication with surrounding vehicles, such as forward collision warnings, emergency braking alerts, and lane change warnings. V2P communication deals primarily with pedestrian safety alerts. V2N communication revolves around intelligent control services such as route planning, remote control, and dynamic map downloads. Among these, V2V refers to the communication between vehicles through onboard terminals, playing a crucial role in V2X[3] by ensuring safer and more enjoyable driving experiences through the exchange of information like speed, location, and direction, allowing for real-time prediction and avoidance of potential collisions, optimizing traffic flow, reducing accidents, and enhancing energy and road use efficiency.

With the global commercial rollout of the fifth-generation mobile network (5G), Cellular-V2X (C-V2X) communication assisted by 5G base station gNodeB (gNB) has been recognized as a promising vehicular communication method due to its low latency[4]. In 5G, security requirements mainly include confidentiality, integrity, authenticity, privacy, and availability. Meanwhile, 5G-V2V is becoming increasingly important as V2X communication involves the exchange of sensitive information among vehicles, infrastructure, and other entities, providing ultra-low latency and excellent reliable connectivity under high mobility and density for situational awareness of the surrounding environment. Therefore, securing 5G-V2V is crucial[5].

Authentication and encryption are fundamental components of secure communication protocols in the 5G-V2X domain[6]. Authentication ensures the verification of the identities of communication entities, preventing unauthorized access and message forgery. On the other hand, encryption ensures the confidentiality of transmitted data, preventing unauthorized disclosure. In direct vehicle communication, it is essential to ensure the authenticity of the transmitter's identity to prevent impersonation during message transmission. Mutual authentication is fundamental to ensure that both parties are communicating with legitimate entities and not attempting to impersonate someone else[7].

However, while information sharing among vehicles offers many benefits, it also brings the possibility of various malicious attacks. To address the security and privacy issues in VANETs, various authentication schemes have been proposed, such as those based on symmetric and public key cryptography[8][9]. In these schemes, a Trust Authority (TA) controls the composition of public and private key pairs and their distribution to legitimate members. Most of these rely on infrastructure like Road Side Units (RSUs) or TAs for centralized management of vehicle information. However, such external facilities may not be available in remote/rural areas[10].

This paper proposes an ECC-based lightweight anonymous V2V mutual authentication protocol, EAIA, which can authenticate vehicles locally without the involvement of trusted authorities or other infrastructure.

1. In the mutual authentication between vehicles process, the proposed protocol achieves authentication and session key negotiation without the involvement of TAs and RSUs, enabling vehicles to communicate in scenarios lacking traffic infrastructure.

2. Using temporary anonymous identities, nodes cannot resolve each other's real identities, thereby protecting privacy. Once the authentication session is initiated, the established temporary session key is used, avoiding the burden of key management and making it difficult for attackers to obtain keys or tamper with messages.

3. Considering the limited computational resources of vehicles, the authentication protocol is designed to be lightweight and efficient while resisting various complex typical attacks.

The remaining paper is organized as follows: Section II presents the literature survey. Section III explains the background knowledge. Section IV shows the system model. Section V details the proposed protocol. Section VI presents the security analysis. Section VII displays the results of the performance evaluation. Section VIII concludes the paper.

## II. RELATED WORK

In this section, we systematically compare and analyze related research on inter-vehicle mutual authentication. Despite the critical importance of mutual authentication for securing vehicle communication, research focused on inter-vehicle mutual authentication in the absence of infrastructure remains limited. Existing studies primarily concentrate on infrastructure-supported authentication mechanisms, which, while providing certain solutions, still exhibit significant weaknesses in terms of security or efficiency.

Firstly, many studies rely on infrastructure support to achieve inter-vehicle mutual authentication. These studies often use Public Key Infrastructure (PKI) or other centralized authentication mechanisms, involving trusted third parties or RSUs for distributing and verifying authentication information. For example, some research proposes methods for distributing digital certificates and keys through RSUs to provide authentication services when vehicles enter a specific area [11]. In this paper, a robust authentication scheme is proposed, using decentralized authentication through vehicle-to-roadside unit communication to continuously update the vehicle sensor network (VSN) information. Only trusted vehicles communicate via a V2V secret communication channel established with RSU assistance [12]. In this scheme, On-Board Units (OBUs) generate anonymous identities and temporary encryption keys to initiate authentication sessions. Secondly, the legitimacy of vehicles' real and anonymous identities can be verified by a TA. When a vehicle wishes to establish a secure session with another vehicle, it first requests authentication from nearby RSUs. After preliminary processing, the RSU forwards the request through a wired channel to higher-level authorities for identity and reputation verification. Finally, with RSU assistance, session keys are negotiated [13]. A lightweight anonymous batch verification protocol for V2I and V2V authentication based on Elliptic Curve Cryptography (ECC) has been proposed, using Physical Unclonable Functions (PUF) and biometric keys to prevent RSU capture attacks and OBU intrusions. The design incorporates a feature embedding strategy with dynamic pseudonyms to restore malicious vehicle identities via a TA. Vehicles within the same RSU domain can authenticate each other and establish session keys, allowing communication in any RSU domain or traffic infrastructure-free environment without repeating authentication.

[14] Proposed an efficient privacy-preserving mutual authentication protocol for secure V2V communication in VANETs, setting up law executors (LEs) assumed to be trustworthy and acting as mobile authentication servers. Initially, ordinary vehicles authenticate solely through LEs. If successfully authenticated, the vehicle becomes trusted. Untrusted OBUs can be authenticated by LEs in the regular authentication phase or by trusted OBUs in the trust extension phase. Trusted vehicles can then communicate securely in the secure communication phase. When the key's lifecycle expires, the vehicle's status becomes untrusted, and key revocation is enforced. [15] Designed a lightweight mutual authentication protocol for vehicular networks using cryptographic operations, proposing the concept of vehicle servers (VS) to provide updated real-time information to requesting vehicles. When a vehicle wishes to communicate with another (e.g., send a warning message), it encrypts the message using its key. The receiving vehicle forwards the requesting vehicle's identity to the VS, which verifies the legitimacy of both vehicles and, if successful, sends the requesting vehicle's key to the other vehicle via a secure channel for decryption and communication. [16] Proposed a hybrid D2D message authentication (HDMA) scheme for 5G-supported VANETs using a novel group signature-based algorithm for V2V mutual authentication. Vehicles entering an RSBS coverage area and seeking services send pseudonyms for initial V2I authentication. If successful, RSBS generates and sends local group signature private key pairs for V2V communication within the same area. [17] Aimed to perform key updates among vehicles without RSU coverage, using secure channels established with existing public group keys. The requesting party proves its identity using an asymmetric key-based challenge-response mechanism to the key-providing party.

While these methods can be effective in certain scenarios, they depend on comprehensive infrastructure deployment and maintenance, which is not feasible in infrastructure-free or weak infrastructure environments.

Secondly, although research on inter-vehicle mutual authentication without infrastructure is limited, some studies propose solutions addressing the authentication issue through various techniques. [18] Proposed a blockchain-based protocol for V2I authentication, V2I handover authentication, and V2V broadcast authentication without relying on traffic infrastructure (e.g., RSU) or trusted authority (TA). Vehicles can broadcast accident reports to other vehicles without RSUs, but pre-broadcast authentication still relies on V2I, not purely V2V authentication. [19] Proposed an identity-based cryptography (IBC) scheme for V2V identity authentication and key agreement. C-V2X devices use their vehicle identification (VID) as their public key. The key management center (KMC) generates private keys for C-V2X devices based on their VID. C-V2X devices transmit secret data encrypted with the recipient's public key and verify each other using an identity-based cryptographic challenge-response protocol, negotiating working keys for encrypted communication. [20] Proposed a new 5G-V2X architecture leveraging network slicing (NS) to ensure different V2X service characteristics and analyzed security requirements based on V2X service types. When vehicles request subscribed services, NSSF determines whether to allow the request, mapping allowed and configured S-NSSAI, and isolating mapped slices. VCF, responsible for V2X service authorization and revocation, transmits service credentials V2V without revealing sensitive vehicle information. Vehicles with V2V service credentials and proximity discover each other, negotiating shared keys for secure communication via 5G millimeter-wave direct V2V links or operating PC5 interfaces. [21] Proposed an authentication mechanism executable in infrastructure-free VANET environments based on LiDAR information for vehicle authentication, called LiDAR. LiDAR authenticates vehicles using available hardware based on shared surrounding information, establishing secure V2V symmetric keys after verifying common object types, distances, and angles.

Most existing methods remain unsuitable for environments lacking RSUs. Moreover, due to their innovative nature, these schemes often do not conform to traditional vehicular network frameworks, resulting in practical application limitations. However, some studies demonstrate feasibility and reference value even without RSUs, leveraging conventional vehicular network frameworks. [22] Proposed a multi-region authentication and privacy protection protocol (MAPP) based on bilinear pairing cryptography and short digital signatures. The protocol supports message and identity authentication within single and multiple regions, enhancing the security and availability of 5G-V2X networks. Although bilinear pairing-based schemes offer robust security, they typically entail high computational costs, making them unsuitable for resource-constrained vehicles. [23] Highlighted the main drawback of pseudonym-based CPPA schemes—frequent pseudonym updates due to link attacks—posing a significant burden on VANETs. To address this, the scheme proposed V2V communication without pseudonyms, avoiding pseudonym update and management issues while supporting secure V2V communication, ensuring only legitimate vehicles access transmitted messages. Some researchers proposed pseudonym authentication schemes for 5G vehicular networks, involving complex and time-consuming operations. [24] Proposed a fog computing-based pseudonym authentication (FC-PA) scheme to reduce performance overhead in 5G vehicular networks, using a single scalar multiplication operation of elliptic curve cryptography to prove information. [25] Addressed the many challenges of secure and efficient mobility management due to frequent handovers and large-scale vehicle communication, proposing a protocol to overcome existing authentication protocol shortcomings. Using Elliptic Curve Diffie-Hellman (ECDH) for secure key transmission, Elliptic Curve Digital Signature Algorithm (ECDSA) for signing message verifiers, timestamps, and random numbers to prevent replay attacks, and a combination of group keys, temporary group keys, tickets, and message authentication codes. [26] Proposed a lightweight authentication scheme for V2V communication, using cryptographic concepts for inter-vehicle message transmission, cross-checking vehicle identities with stored TPD values, and ensuring only vehicles with proven authenticity participate in the message transmission phase. [27] Introduced a new certificate-less AKE protocol for V2V communication in vehicular networks, achieving partial non-repudiation by computing partial keys by users and resisting temporary key leakage attacks by merging temporary secrets and private keys into shared information.

These studies highlight the importance of designing a lightweight, efficient, and secure direct inter-vehicle mutual authentication mechanism. An ideal solution should effectively handle authentication and data security issues without relying on external infrastructure while ensuring efficient communication remains unaffected. Our research aims to fill this gap by proposing a new authentication framework to address the limitations of existing methods in infrastructure-free environments.

## III. PRELIMINARIES

Before introducing the proposed EAIA scheme for V2V mutual authentication, we briefly describe some fundamental concepts and technical preliminaries used in the protocol design. Section III-A covers the relevant basics of elliptic curve cryptography. Section III-B outlines several mathematical problems utilized in the establishment and proof of the proposed scheme. Lastly, Section III-C introduces certificate-less public key cryptography.

### A. Elliptic Curve Cryptosystem

Elliptic Curve Cryptography (ECC) is a public key encryption algorithm based on the mathematics of elliptic curves. It uses points on an elliptic curve and their operations (such as addition and scalar multiplication) to construct encryption and decryption mechanisms. The general form of the elliptic curve equation is $y^2 = x^3 + ax + b$, where $a$ and $b$ are constants. Specifically, let $q > 3$ be a prime number, and define an elliptic curve over the finite prime field $Z_q = \{0, 1, \ldots, q-1\}$ as the set of solutions $E_q(a, b)$ to the congruence $y^2 \equiv x^3 + ax + b \pmod{q}$, where $a, b \in Z_q$ are constants satisfying the non-singularity condition $4a^3 + 27b^2 \not\equiv 0 \pmod{q}$.

Compared to traditional RSA algorithms, ECC can achieve the same level of security with shorter key lengths, significantly improving computational efficiency and reducing storage requirements. This advantage makes ECC particularly suitable for resource-constrained environments.

*B. Mathematical Problems*

ECC is based on the difficulty of the Elliptic Curve Discrete Logarithm Problem (ECDLP), which involves computing the inverse of point multiplication on an elliptic curve (i.e., the discrete logarithm). Compared to traditional RSA encryption, ECC can achieve the same level of security with shorter keys, making it widely adopted in modern cryptography. It is used in digital signatures (e.g., ECDSA), key exchange (e.g., ECDH), and data encryption, forming a crucial part of SSL/TLS, blockchain technology, and various encryption communication protocols. The proposed V2V mutual authentication protocol, EAIA, mainly utilizes ECDLP and ECDH.

1）Elliptic Curve Discrete Logarithm Problem (ECDLP)

The ECDLP is the foundation of elliptic curve cryptography. Given an elliptic curve $E$ and two points $P$ and $Q$ on it, if there exists an integer $k$ such that $Q = kP$, then $k$ is the discrete logarithm of $Q$ with respect to $P$. Computing this $k$ is known as solving the ECDLP. This problem is considered very difficult, especially over large prime fields, providing the security basis for elliptic curve cryptography.

2）Elliptic Curve Diffie-Hellman (ECDH)

ECDH is a key exchange protocol based on elliptic curves. It allows two participants, who do not share any prior secret information, to agree on a shared key over an insecure communication channel. The steps are as follows:

a. Key Generation: Each participant chooses a private key $a$ and $b$, and computes the corresponding public keys $A = aP$ and $B = bP$, where $P$ is a generator on the elliptic curve.

b. Key Exchange: Participants exchange their public keys.

c. Shared Key Computation: Each participant uses the other's public key and their own private key to compute the shared key. The first participant computes $S = aB$, and the second participant computes $S' = bA$. Since $S = S' = abP$, both participants arrive at the same shared key.

*C. Certificate-less Public Key Cryptography (CL-PKC):*

CL-PKC is a novel public key cryptographic system introduced by Al-Riyami and Paterson in 2003. It was designed to address the shortcomings of traditional public key cryptography and identity-based public key cryptography (ID-PKC).

In the CL-PKC system, a Key Generation Center (KGC) generates a partial private key for the user based on their identity information. The user then selects a random value and combines it with the partial private key to generate a complete private key. The user's complete private key consists of two independent secret components: one is the partial private key derived from the KGC based on the user's identity, and the other is the random key generated by the user. These two secret components are independent of each other, meaning that one cannot be used to compute the other.

Consequently, the KGC cannot determine the user's complete private key, nor can the user deduce the partial private key generated by the KGC.

The implementation of CL-PKC involves the following seven algorithms: System Initialization (Setup), Partial Private Key Extraction (Partial-Private-Key Extract), Secret Value Setting (Set-Secret-Value), Private Key Generation (Set-Private-Key), Public Key Generation (Set-Public-Key), Encryption (Encrypt), and Decryption (Decrypt). In practical applications, these algorithms work together to provide efficient and secure encryption, signing, and key exchange functionalities.

Compared to traditional Public Key Infrastructure (PKI), CL-PKC simplifies public key management, reduces reliance on centralized authorities, and enhances system flexibility and security. Therefore, CL-PKC is particularly well-suited for the scenarios we have discussed.

## IV. SYSTEM MODEL

*A. System Model*

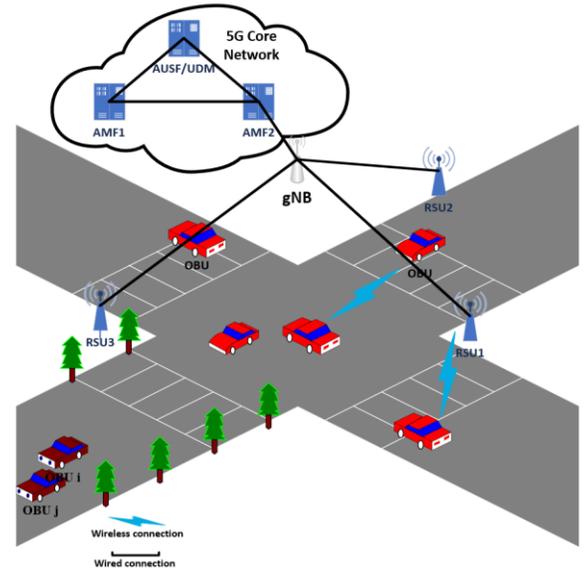

**Fig. 1.** System Model.

As shown in Figure 1, the 3GPP 5G system primarily consists of the 5G Core Network (5GC) and the 5G Radio Access Network (RAN). In the 5G RAN, each vehicle and RSU connect to the network via wireless channels through gNB. The gNB connects to the 5GC via wired channels. The system architecture of the 5G core network includes multiple functions such as Access and Mobility Management Function (AMF), Authentication Server Function (AUSF), and Unified Data Management (UDM). During the registration and authentication process, the AMF handles access requests, authentication, registration, mobility management, and handover control for user equipment (UE). It maintains the authentication and key management for vehicles and interacts with the AUSF. The AUSF is responsible for the authentication and authorization of user equipment, ensuring secure network access and functioning as an authentication server. The UDM provides storage and management of user data, supporting user profiles, authentication data, and subscription data.

The 5G-supported V2V identity authentication protocol involves four participants: OBU, (RSU), gNB, and the 5G Core Network. Except for the OBU, the other participants are involved only in the registration phase. In the authentication phase, only two OBUs participate.

On-Board Unit (OBU): In the 5G-V2X (Vehicle-to-Everything) ecosystem, the OBU is a device installed in vehicles responsible for communication with other vehicles and infrastructure. It acts as a vehicle controller with computational capabilities, sending and storing traffic information and exchanging road condition data between vehicles. Each vehicle has a unique real identity, denoted as RID, pre-installed by the manufacturer, and serves as a unique registration identifier.

Road Side Unit (RSU): The RSU is a critical infrastructure component in the 5G-V2X network, serving as a bridge between vehicles and the network infrastructure. Strategically placed along roads, RSUs provide connectivity and facilitate data exchange between vehicles and the network. They support direct V2X communication and can connect to gNB via fiber optic or high-speed Ethernet to relay information across the core network. RSUs enhance road safety and traffic management by enabling low-latency and high-reliability communication. However, RSUs are not entirely trustworthy entities and are susceptible to attacks by malicious adversaries.

5G Base Station (gNB): The next-generation NodeB (gNB) is a 5G base station providing Radio Access Network (RAN) connectivity within the 5G-V2X framework. The gNB manages radio resource allocation and communication connections in the cellular network. By communicating with vehicles and RSUs, the gNB offers wide-area coverage and supports high mobility. It is also responsible for connecting to the core network to ensure data transmission and processing.

5G Core Network (5GC): The 5G Core Network (5GC) is the backbone of the 5G-V2X ecosystem, coordinating various network functions and managing data flow between the access network and external services. It comprises multiple Network Functions (NFs), including the Access and Mobility Management Function (AMF), Session Management Function (SMF), User Plane Function (UPF), and Policy Control Function (PCF), among others. Each entity performs specific network functions to ensure the efficient operation of the 5G network.

*B. Attack Model*

In our research, the network attack model is the Dolev-Yao model, which is a widely used attack model in the field of information security for analyzing and verifying the security of communication protocols. This model assumes that the attacker has full control over the network, enabling them to intercept, modify, replay, and inject messages. The strength of the Dolev-Yao model lies in its provision of extensive capabilities to the attacker, thus any protocol that can ensure security under this model is considered to be highly secure.

In the context of 5G vehicular networks, the Dolev-Yao model is employed to evaluate and enhance the security of communication protocols. For instance, in communications between vehicles and infrastructure as well as between vehicles themselves, attackers might attempt to intercept and alter data to carry out man-in-the-middle attacks or identity forgery. Specifically, the following scenarios can be expressed:

1. Eavesdropping, Interception, Modification, or Deletion of Messages: An adversary can eavesdrop on, intercept, modify, or delete publicly transmitted messages, potentially compromising the confidentiality and integrity of the communication. Passive attacks may also be used to gather sensitive information.

2. Replay Attacks: An adversary might capture and resend previously transmitted data packets to deceive other vehicles or infrastructure, leading to the propagation of duplicate or misleading information.

3. Message or Credential Forgery: An adversary might forge false messages or credentials to impersonate legitimate vehicles or road infrastructure, causing the dissemination of incorrect traffic information or false warning messages, thereby impacting road safety.

4. Man-in-the-Middle Attacks: An adversary could position themselves between communicating parties, establishing normal connections with both sides and deceiving them into exchanging data through the attacker.

Meanwhile, it is assumed that the connections between the 5G core network and gNB, as well as between gNB and RSU, are secure due to their wired nature. Additionally, the connections between vehicles and RSUs or gNBs are assumed to be insecure. Furthermore, all network functions within the 5G core network are assumed to be trusted since they are located in secure environments (such as hardware security modules).

V. PROPOSED SCHEME

In this section, we delineate the EAIA scheme, which leverages certificateless cryptographic techniques for anonymous authentication and key agreement. The proposed protocol is structured into five distinct phases: 1) Initialization phase to generate foundational parameters, 2) Registration to enroll OBUs into the network, 3) V2I mutual authentication to ensure secure of V2I communication, 4) Mutual Authentication between OBUs to ensure secure and verified V2V communication with a session key agreement, and 5) Pseudonym Update for periodic anonymity renewal. The protocol employs a set of specific symbols, each with a defined purpose and function, as detailed in Table I.

Identify applicable funding agency here. If none, delete this text box.

TABLE I Notation and Description of EAIA

| Notion | Description |
|---|---|
| $\lambda$ | The security parameter for the system |
| p,q | The large primes. |
| $l$ | The length of session keys |
| G | The additive group over the elliptic curve cryptography |
| P | The generator of G. |
| s | The master key of system. |
| $P_{pub}$ | The public key of system |
| $h_i$ | The hash functions(i=1,...,4) |
| a,b,d | The random number chosen in the |
| $T_a, T_b$ | The timestamps chosen in the communication |
| $RID_i, RID_j$ | The real identity of $Vehicle_{i/j}$ |
| $ID_i, ID_j$ | The pseudo identity of $Vehicle_{i/j}$ |
| $(x_{i/j}, y_{i/j})$ | The static private key pair of $Vehicle_{i/j}$ |
| $(X_{i/j}, Y_{i/j})$ | The static public key of $Vehicle_{i/j}$ |
| $(SK_{i-j}, SK_{j-i})$ | The session keys generated in proposed |

### A. System Initialization

At this stage, the AMF selects and generates the system's public parameters. Given $\lambda$ as the system's security parameter, and $|q|$ and $p$ as large prime numbers, the AMF selects an appropriate additive subgroup $G$ based on the elliptic curve $E$ over the finite field $F_p$. The subgroup $G$ is generated by $P$ and has an order denoted as $q$. Subsequently, the AMF selects its master key $s$ and computes the corresponding AMF public key $P_{pub} = sP$. Additionally, the AMF defines four secure one-way hash functions, denoted as $h_1:\{0,1\} \to Z_q^*$, $h_2: G \to \{0,1\}^*$, $h_3:\{0,1\}^* \to \{0,1\}^l$, and $h_4:\{0,1\}^l \to \{0,1\}^\lambda$. The final outcome of this process is the dissemination of the selected public parameters $(G, P, P_{pub}, h_1, h_2, h_3, h_4)$ throughout the system.

### B. Registration

To protect user privacy, each vehicle must first register with the AMF to obtain its pseudonym before participating in the communication. Since the connection between RSUs and OBUs is wireless, it is inherently open and dynamic, rendering the channel insecure. Therefore, to register vehicles onto the 5G network, our scheme employs the 5G Authentication and Key Agreement (5G-AKA) mechanism introduced by the 3GPP standard to ensure compatibility with the 5G network. This creates a secure channel between the OBU and the AMF, through which the registration process is securely completed. A secure channel ensures the confidentiality and integrity of the data transmitted through it. The detailed process is described below and illustrated in Fig.2.

*Step 1:* First, $OBU_i$ sends its registration request and registration information to the AMF through the abovementioned secure channel. The information includes the distinct vehicle identity, denoted by the unique vehicle identifier $RID_i$, essentially the chassis number of the vehicle. The chassis number is more accurately called the "vehicle identification number", commonly known as the frame number, which is the unique certificate for each vehicle. Concurrently, the registration process involves $OBU_i$ generating a random number $x_i$ within the group $Z_q$. Subsequently, the registration request, specifically, { $RID_i, X_i$}, where $X_i=x_iP$, is sent to the AMF.

*Step 2:* After AMF receives the registration request { $RID_i, X_i$}, it verifies the legality and uniqueness of the identity and selects $r_i \in Z_q$, and then calculates $R_i=r_iP$, $H_i=h_1(ID_i||X_i||R_i)$, $y_i=(r_i+sH_i)$, $Y_i=R_i+P_{pub}H_i$. At the same time, the pseudonym of the OBU $ID_i=RID_i \oplus h_1(s||R_i)$ is generated. After completing these calculations, send ($y_i$, $R_i$, $Y_i$, $ID_i$) to the $OBU_i$ through the secure channel.

*Step 3:* After obtaining ($y_i$, $R_i$, $Y_i$, $ID_i$), $OBU_i$ checks whether $y_iP = R_i+P_{pub}H_i$ is held. If not, abort it. Otherwise, $OBU_i$ will store the secret private key $(x_i+y_i)$ and publish the public key $(X_i+Y_i)$. Then, the registration of $OBU_i$ is completed.

*Step 4:* The registration steps for $OBU_j$ are the same as those for $OBU_i$. $OBU_j$ will store the secret private key $(x_j+y_j)$ and publish the public key $(X_j+Y_j)$. Then, the registration of $OBU_j$ is completed.

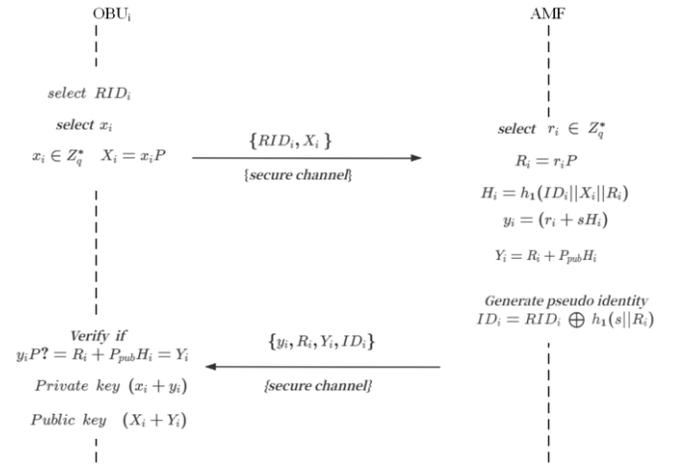

**Fig. 2.** Registration

### C. Mutual Authentication

After completing the registration process, mutual authentication between OBUs is required to initiate the establishment of a session key with the nearest proximal OBU. The procedure assumes that $OBU_j$ sends an authentication request to $OBU_i$ and wishes to establish a session key. This procedure is delineated in the following steps and is represented in Fig.3.

*Step 1:* If $OBU_j$ requests authentication with $OBU_i$, $OBU_j$ sends its own ID $ID_j$, public key $(X_j+Y_j)$, and an authentication request to $OBU_i$. Once $OBU_i$ receives the request from $OBU_j$, it generates random values of $a, b \in Z_q$ and calculates $A= aP$, $B= bP$, $M = b(X_j+Y_j)$, $N= h_2(M)\oplus(ID_i||A)$. This means that $OBU_i$ encrypts its identity using $OBU_j$'s public key, so only $OBU_j$ can extract $OBU_i$'s identity. Then $OBU_i$ calculates $\sigma = b^{-1}(h_1(ID_i||A||T_a)+y_i)$ as its signature where $T_a$ represents the current timestamp. Then $OBU_i$ sends {$B, N, \sigma, T_a$} to $OBU_j$.

*Step 2:* After receiving the message {$B, N, \sigma, T_a$} from $OBU_i$, if $T_a$ is not fresh, $OBU_j$ aborts this session. Otherwise, $OBU_j$ computes $M'=(x_j+y_j)B$ and derives $ID_i'||A' = h_2(M')\oplus N$. Since we can get $M=M'$ from (1), from $N= h_2(M)\oplus(ID_i||A)$ and $ID_i'||A'= h_2(M')\oplus N$, we can get $ID_i||A= ID_i'||A'$. In this way, it is guaranteed that what $OBU_i$ sends to $OBU_j$ can be received by $OBU_j$. Complete identity authentication for $OBU_j$.

$$\begin{aligned} M' &= (x_j + y_j)B \\ &= b(x_j + y_j)P \\ &= b(X_j + Y_j) \\ &= M \end{aligned} \quad (1)$$

Then, $OBU_j$ verifies the validity of $OBU_i$ by verifying the signature. $OBU_j$ checks whether $\sigma B = h_1(ID_i'||A'||T_a)P + Y_i$ holds. If it is held, the identity authentication is achieved. The correctness of identity authentication can proved by (2). By doing this, it is guaranteed that the sender is $OBU_i$. Complete identity authentication for $OBU_i$.

$$\begin{aligned} \sigma B &= (b^{-1}(h_1(ID_i||A||T_a) + y_i))B \\ &= b(b^{-1}(h_1(ID_i||A||T_a) + y_i))P \\ &= h_1(ID_i||A||T_a)P + y_iP \\ &= h_1(ID_i||A||T_a)P + Y_i (2) \end{aligned}$$

If the $OBU_i$ verification is successful, it implies that the identity authentication of both parties is successful. Then the session key negotiation begins. $OBU_j$ randomly selects $D=dP$, $d \in Z_q$, and calculates $k=h_1(ID_i'||ID_j||A'||D||T_b)$, $SK_{j-i}=h_3(ID_i'||ID_j||d(kA'+X_i))$, $\eta=h_4(ID_i'||ID_j'||SK_{j-i}||T_b)$. Then, $OBU_j$ sends $\{D, \eta, T_b\}$ to $OBU_i$.

*Step 3:* After receiving the response of $OBU_j$, if $T_b$ is not fresh, $OBU_i$ aborts this session. Otherwise, $OBU_i$ computes $k=h_1(ID_i||ID_j||A||D||T_b)$ and $SK_{i-j}=h_3(ID_i||ID_j||D(ka+x_i))$. Finally, it verifies whether $\eta=h_4(ID_i||ID_j||SK_{i-j}||T_b)$ is successful. If the equation holds, mutual authentication is completed with the session key successfully negotiated. Therefore, we have $SK_{i-j} = SK_{j-i}$ because the following equation (3) holds:
$$D(ka + x_i) = d(ka + x_i)P = d(kA + X_i) \quad (3)$$

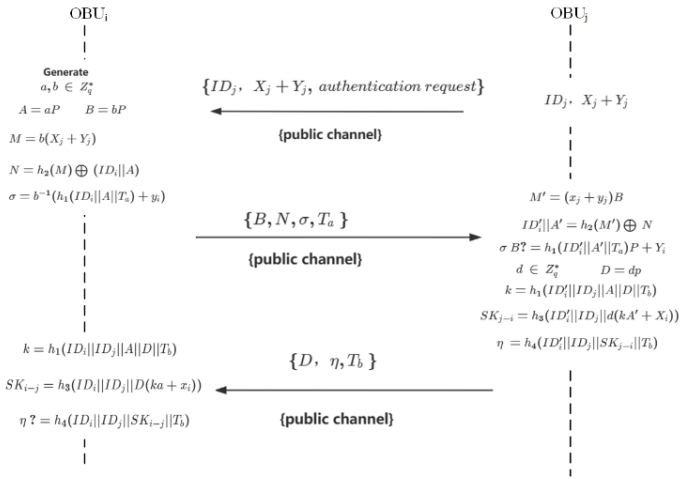

**Fig. 3.** Mutual Authentication

### D. Pseudo-identity Update

Regularly changing the pseudonyms of vehicles can effectively prevent vehicles from being tracked and personal information leaked. Pseudonymous updates protect the privacy of the vehicles while allowing the vehicles to continue participating in necessary communications and data exchanges without revealing their true identities.

*Step 1:* The validity period of each pseudonym is $\Delta t$. When $|T-T_b| \geq \Delta t$, a pseudonym update is required, where $T$ represents the current time, $\Delta t$ is the period of the system parameter, and $T_b$ is approximately equal to the certification end time. Then, the OBU at the vehicle sends $\{\sigma, ID_i A, R_i\}$ as the pseudonym update request to the AMF.

*Step 2:* Upon receipt of the request, the AMF verifies the legitimacy of the identity by calculating $\sigma B = h_1(ID_i||A||T)P + Y_i$. After the validation, the TA embarks on the calculation of $RID_i = ID_i \oplus h_1(s||R_i)$, and simultaneously selects a new random variable, referred to as $r'_i \in Z_q$, by which, the TA proceeds to compute and generate a new pseudonym, identified as $ID_i' = RID_i \oplus h_1(s||R_i')$. In the subsequent steps, the TA performs the computation of $Q = h_1(y_i) \oplus ID_i'$ culminating in the dispatch of $(Q, T_c)$ to the vehicle. $T_c$ represents the current timestamp .By hashing the ID with the $y_i$, it avoids the transmission of the ID in plaintext and the illegal user obtaining the ID.

*Step 3:* After receiving the message $(Q, T_c)$, if $T_c$ is not fresh, $OBU_i$ aborts this session. Otherwise, it then computes to generate the new pseudonym $ID'_i = Q \oplus h_1(y_i)$.

## VI. SECURITY EVALUATION

In this section, we first conduct a qualitative security analysis of the proposed protocol. This is followed by a logical correctness proof, where we evaluate the EAIA scheme using BAN logic. Finally, we perform formal verification testing with the Scyther tool. The results indicate that the proposed scheme meets the security objectives.

### A. Security Analysis

In this section, we will discuss and analyze the security properties of the main component of our proposed EAIA protocol, which is the mutual authentication between vehicles.

*1) Mutual Authentication:* For two vehicles requiring mutual authentication, $OBU_i$ encrypts its identity using $OBU_j$'s public key, ensuring that only $OBU_j$ can extract $OBU_i$'s identity. To validate the signature and verify the validity of $OBU_i$, $OBU_j$ uses its private key to compute $M'$. Only the target vehicle $OBU_j$ can obtain the correct identity and random number $A$ of $OBU_i$ and verify $OBU_i$'s identity. Therefore, $OBU_i$ can also validate the validity of $OBU_j$.

*2) Secure Session Key Agreement:* The integrity and confidentiality of the session key are ensured by the principles of ECCDH. If an adversary could forge a session key between vehicles, it would imply that the adversary could solve the Computational Diffie-Hellman problem, which is known to be difficult as discussed in Section III-B.

*3) Anonymity and Privacy Protection:* In the proposed protocol, real identities are anonymized by generating pseudonyms, which are then obscured by hash functions.

During the authentication process, the messages $\{B, N, \sigma, T_a\}$ and $\{D, \eta, T_b\}$ do not directly transmit identities. Instead, identities are linked with random numbers, preventing the sender's identity from being disclosed. Thus, the proposed protocol satisfies user anonymity and privacy protection requirements.

4) Resistance to Man-in-the-Middle Attacks: During authentication, authenticated OBUs can verify the requesting OBU by generating its signature σ using the requesting vehicle's private key $(x_i + y_i)$. Without this private key, the authenticated vehicle cannot extract the identity of the requesting vehicle. If an adversary attempts a man-in-the-middle attack, they must possess $(x_i + y_i)$ to complete identity authentication and key exchange. However, $(x_i + y_i)$ remains unknown to any adversary.

5) Resistance to Impersonation Attacks: To successfully impersonate a vehicle, an attacker needs to know the private key $(x_i + y_i)$ of the requesting OBU to generate a legitimate signature σ. Since the attacker cannot obtain the private key $(x_i + y_i)$ of the requesting vehicle, they cannot generate a valid signature, thus preventing impersonation.

6) Resistance to Replay Attacks: In the proposed protocol, timestamps $T_a$ and $T_b$ ensure that session keys cannot be reused by adversaries to disrupt the mutual authentication process.

7) Perfect Forward Secrecy: Forward secrecy ensures that even if participants' long-term private keys and previous session keys are compromised, the current session key remains secure. In the proposed protocol, if the private keys $(x_i + y_i)$ and $(x_j + y_j)$ of two participants are leaked, adversaries still cannot access the session key without knowing the temporary keys. Assuming the current session keys are $SK_{j-i} = h_3\left(ID_i' \|ID_j\| d(kA' + X_i)\right)$ and $SK_{i-j} = h_3\left(ID_i \|ID_j\| D(ka + X_i)\right)$, adversaries might access $ID_i$ and $ID_j$. For $d(kA' + X_i)$, it should be computed as $d(ka + x_i)P$, where $k = h_1(ID_i\|ID_j \| A \| D \| T_b)$. However, adversaries do not have access to the temporary keys $d$ or $a$ involved in generating the session key. Therefore, our protocol meets the requirements for forward secrecy.

8) Resistance to Random Number Leakage: During the authentication process, even if the random numbers $a, b$, and $(d)$ are leaked, adversaries cannot generate the correct session key because the secret keys of the vehicles are used as part of the session key.

### B. Formal Proof by BAN Logic

Burrows-Abadi-Needham Logic (BAN Logic), initially proposed by Burrows et al., is a formal method used to analyze and verify the security properties of authentication protocols. It has been employed to formally derive the logical correctness of security protocols. BAN-logic provides a logical framework to describe the beliefs of trusted parties involved in an authentication protocol and the evolution of these beliefs.

To perform verification, the protocol must first be translated into an idealized version. Assumptions and goals need to be stated. Then, derivation rules are manually applied to achieve the goals. The key elements of BAN-logic include principals (entities such as users or servers), messages, and the beliefs, goals, and trust relationships of these principals. The symbols used in BAN-logic are listed in Table II.

TABLE II Notation and Meaning of Ban-Logic

| Notation | Meaning |
| --- | --- |
| $P \mid \equiv X$ | $P$ believes the message $X$ |
| $P \triangleleft X$ | $P$ sees the message $X$ |
| $P \mid \sim X$ | $P$ said the message $X$ |
| $P \Rightarrow X$ | $P$ has authority on $X$ |
| $\#(X)$ | $X$ is fresh |
| $\langle X \rangle_K$ | $X$ is combined with a secret $K$ |
| $\{X\}_K$ | $X$ is encrypted under the key $K$ |
| $P \stackrel{K}{\leftrightarrow} Q$ | $K$ is a secret key shared between $P$ and $Q$ |
| $P \stackrel{K}{\rightleftharpoons} Q$ | $K$ is a shared secret between $P$ and $Q$ |
| $\stackrel{K}{\mapsto} P$ | $P$ has a public key $K$ corresponding to a private key $K^{-1}$. |

The derivation rules of BAN logic consist of 19 logical rules. Here we only list the rules used in this paper:

Rule 1: Message Meaning Rule

$$\frac{P\mid \stackrel{K}{\equiv} Q, P \triangleleft X_{K^{-1}}}{P\mid \equiv Q\mid \sim X}$$

if P believes that the public key for user Q is K, and P can see the message X signed by the private key of Q is $K^{-1}$, then P believes the message X is sent by Q.

Rule 2: Nonce Verification Rule

$$\frac{P\mid \equiv \#(X), P\mid \equiv Q\mid \sim X}{P\mid \equiv Q\mid \equiv X}$$

This rule means if P believes X is fresh and P believes Q sent X, then P believes Q believes X.

Rule 3: Jurisdiction Rule:

$$\frac{P\mid \equiv Q \Rightarrow X, P\mid \equiv Q\mid \equiv X}{P\mid \equiv X}$$

This rule means if P believes Q has jurisdiction on message X, and P believes Q believes X, then P believes X.

Rule 4: Freshness Rule:

$$\frac{P\mid \equiv \#(X)}{P\mid \equiv \#(X,Y)}$$

This rule means that if part of the formula is fresh, the entire formula is also fresh. It only makes sense to appear in ciphertext. If a part of the ciphertext is fresh, then the entire part is also fresh.

Rule 5: Belief Rule:

$$\frac{P\mid \equiv Q\mid \equiv (X,Y)}{P\mid \equiv Q\mid \equiv X}$$

This rule reflects the consistency of beliefs in different operations of concatenation and segmentation of messages and the transitivity of beliefs in such operations.

The main steps for formal verification using BAN logic are as follows: 1) Idealization: Convert the actual protocol messages into an abstract form that captures the essential security properties. 2) Assumptions: Define the initial beliefs and assumptions about the principals and messages. 3) Derivations: Apply BAN logic derivation rules to deduce new beliefs from the initial assumptions. 4)Verification: Check whether the derived beliefs satisfy the required security properties, such as mutual authentication or key agreement.

1)The Goals of EAIA:

First, we establish our goals. In our model, the AMF is a trusted entity for both $OBU_i$ and $OBU_j$. Therefore, if $OBU_i$ and $OBU_j$ share the same session key with the AMF, we think that $OBU_i$ and $OBU_j$ share a session key. Additionally, the session key $SK$ shared between $OBU_i$ and $OBU_j$ is generated using the random numbers $a$ and $d$. Finally, for simplicity in some parts of the expression, we denote the AMF public key, which is also the system public key, as $pk$. The public keys of OBUi and OBUj, $X_i + Y_i$ and $X_j + Y_j$, are denoted as $ppk_i$ and $ppk_j$ respectively in this section. Hence, the primary goals of our protocol are as follows:

Goal 1: $OBU_i| \equiv (Y_i, y_i)$
Goal 2: $OBU_i| \equiv OBU_i \xleftrightarrow{SK} OBU_j$
Goal 3: $OBU_i| \equiv OBU_j| \equiv OBU_i \xleftrightarrow{SK} OBU_j$
Goal 4: $OBU_j| \equiv OBU_i \xleftrightarrow{SK} OBU_j$
Goal 5: $OBU_j| \equiv OBU_i| \equiv OBU_i \xleftrightarrow{SK} OBU_j$

2) The Idealization of EAIA: For the purpose of formal analysis, the messages exchanged between $OBU_i$ and $OBU_j$ are idealized as follows:

Message 1: $OBU_i \to AMF: RID, X_i$
Message 2: $AMF \to OBU_i: y_i, Y_i, ID, R$
Message 3: $OBU_i \to OBU_j: B, N, \sigma, T_a$
Message 4: $OBU_j \to OBU_i: D, \eta, T_b$

3)The Assumptions of EAIA: Additionally, based on the description of our protocol, we have the following assumptions.

$A_1: OBU_i| \equiv \xrightarrow{pk} AMF$
$A_2: OBU_i| \equiv \#(X_i, Y_i)$
$A_3: OBU_i| \equiv AMF \Rightarrow (X_i, Y_i)$
$A_4: OBU_j| \equiv \xrightarrow{X_i+Y_i} OBU_i$
$A_5: OBU_i| \equiv \xrightarrow{X_j+Y_j} OBU_j$
$A_6: OBU_j|\equiv\#(T_a), OBU_j|\equiv\#(T_b)$
$A_7: OBU_i|\equiv\#(T_a), OBU_i|\equiv\#(T_b)$
$A_8: OBU_j| \equiv OBU_i \Rightarrow OBU_i \xleftrightarrow{SK} OBU_j$
$A_9: OBU_i| \equiv OBU_j \Rightarrow OBU_i \xleftrightarrow{SK} OBU_j$

4）Security Verification of EAIA: Then, we prove that our protocol achieves the goal according to BAN logic, described as follows:

According to the Message 2, we obtain
$OBU_i \triangleleft \{ID_i, Y_i, y_i\}_{pk^{-1}}$ (S1)
By $A_1$ and S1, we employ the Rule 1 to derive
$OBU_i|\equiv AMF|\sim(ID_i, Y_i, y_i)$ (S2)
By $A_2$, we apply the Rule 4 to deduce
$OBU_i| \equiv \#(ID_i, Y_i, y_i)$ (S3)
By S2 and S3, we apply the Rule 2 to derive
$OBU_i|\equiv AMF| \equiv (ID_i, Y_i, y_i)$ (S4)
By S4, we employ the Rule 5 to derive

$OBU_i|\equiv AMF| \equiv (Y_i, y_i)$ (S5)
By $A_3$ and S5, we employ the Rule 3 to deduce
$OBU_i| \equiv (Y_i, y_i)$ (S6)
which satisfies the **Goal 1.**
According to the Message 3, we obtain
$OBU_j \triangleleft \{B, N, \sigma, T_a, OBU_i \xleftrightarrow{SK} OBU_j\}_{ppk_i^{-1}}$ (S7)
By $A_4$ and S7, we apply the Rule 1 to deduce
$OBU_j| \equiv OBU_i|\sim(B, N, \sigma, T_a, OBU_i \xleftrightarrow{SK} OBU_j)$ (S8)
By $A_6$, we employ the Rule 4 to derive
$OBU_j |\equiv \#(B, N, \sigma, T_a, OBU_i \xleftrightarrow{SK} OBU_j)$ (S9)
By S8 and S9, we apply the Rule 2 to derive
$OBU_j| \equiv OBU_i| \equiv ((B, N, \sigma, T_a, OBU_i \xleftrightarrow{SK} OBU_j)$ (S10)
By S10, we employ the Rule 6 to deduce
$OBU_j| \equiv OBU_i| \equiv (OBU_i \xleftrightarrow{SK} OBU_j)$ (S11)
which satisfies the **Goal 5.**
By $A_8$ and S11, we apply the Rule 3 to derive
$OBU_j |\equiv (OBU_i \xleftrightarrow{SK} OBU_j)$ (S12)
which satisfies the **Goal 4.**
According to the Message 4, we obtain
$OBU_i \triangleleft \{D, \eta, T_b,, OBU_i \xleftrightarrow{SK} OBU_j\}_{ppk_j^{-1}}$ (S13)
By $A_5$ and S13, we apply the Rule 1 to derive
$OBU_i| \equiv OBU_j|\sim(D, \eta, T_b, OBU_i \xleftrightarrow{SK} OBU_j)$ (S14)
By $A_7$, we employ the Rule 4 to deduce
$OBU_i |\equiv \#(D, \eta, T_b, OBU_i \xleftrightarrow{SK} OBU_j)$ (S15)
By S14 and S15, we apply the Rule 2 to derive
$OBU_i| \equiv OBU_j| \equiv (D, \eta, T_b, OBU_i \xleftrightarrow{SK} OBU_j)$ (S16)
By S16, we employ the Rule 6 to derive
$OBU_i| \equiv OBU_j| \equiv (OBU_i \xleftrightarrow{SK} OBU_j)$ (S17)
which satisfies the **Goal 3.**
Finally, by $A_9$ and S17, we apply the Rule 3 to deduce
$OBU_i |\equiv (OBU_i \xleftrightarrow{SK} OBU_j)$ (S18)
which satisfies the **Goal 2.**

In summary, all safety objectives were achieved. This means that key agreement and mutual authentication guarantees are achieved. Vehicles are able to establish a secure session key and use it to encrypt messages between them.

*C. Formal Verification by Scyther Tool*

Scyther was first introduced and utilized by Cremers in [28]. It is a powerful formal verification tool used for falsification, verification, and detecting potential attacks, playing a critical role in the analysis of security protocols. The major outstanding feature of Scyther is automatic verification. Scyther can automatically generate and verify all possible execution paths of security protocols, identifying potential security vulnerabilities. Scyther is efficient to complete the verification of complex protocols within a short time. And Scyther uses formal methods to describe protocols, ensuring a high level of precision and rigor in the analysis process. Scyther includes various assertions, such as Non-injective

Synchronization (Nisynch), Non-injective Agreement (Niagree), Aliveness (Alive), and Weak Agreement (Weakagree). Specifically, the term "Niagree" ensures the consistency of message content between sending and receiving, thus guaranteeing message integrity. "Nisynch" refers to the overall mirroring of protocol events, enhancing protocol reliability.

Our study focuses on the key security attributes of mutual identity authentication and secure session key negotiation. In the proposed protocol model, there are two roles: vehicle$_i$ and vehicle$_j$. As shown in Fig.4, both roles achieve synchronization (Nisynch), agreement (Niagree), aliveness (Alive), and weak agreement (Weakagree). Additionally, we verified the security of random numbers $a$, $b$, and $d$. Throughout the entire process, the confidentiality of all encryption keys is maintained. These findings collectively demonstrate the strong security of the proposed protocol. Moreover, the protocol ensures the accessibility of the involved parties during communication and maintains the security of session keys within the established security framework.

In conclusion, the results of the formal verification indicate that the proposed group handover authentication protocol is secure.

**Fig. 4.** Scyther Verification Result of the EAIA

VII. PERFORMANCE EVALUATION

In this section, we evaluate the performance of the proposed EAIA protocol by comparing it with other protocols, specifically those outlined in : HDMA: Hybrid D2D Message Authentication Scheme for 5G-Enabled VANETs [16], An Efficient Privacy-Preserving Mutual Authentication Scheme for Secure V2V Communication in Vehicular Ad Hoc Network [14], Efficient Privacy-Preserving Dual Authentication and Key Agreement Scheme for Secure V2V Communications in an IoV Paradigm [12], Secure Privacy-Preserving V2V Communication in 5G-V2X Supporting Network Slicing [20]. These protocols are referred to as HDMA PPMA, PPDAS, and SPPC. To accurately assess performance, we set the length and time cost of each parameter during simulation. Specifically, we define |ID| as the length of each participant's identity, |T| as the length of a timestamp, |G| as the length of an element in the cyclic group G, |H| as the length of a hash function output, and |q| as the length of an element in $Z_q^*$. The lengths are set as follows: |G| = 320 bits, $|Z_q^*|$ = 160 bits, |ID| = 256 bits, |H| = 512 bits, and |T| = 32 bits.

TABLE III  Comparison of Computational Cost

| Notations | Description | OBU Computation Time(μs) |
|---|---|---|
| $T_{hash}$ | Hash(SHA-256) | 2 |
| $T_{sm}$ | Scale multiplication | 576 |
| $T_{pa}$ | Point addition related to the ECC | 20 |
| $T_{me}$ | Modular exponential operation (1024 bits) | 249 |
| $T_{enc}/T_{dec}$ | AES-256 encryption/ decryption | 530/7425 |
| $T_{sig}$ | The computation time of an ECDSA signature generation based on the secp256k1 curve | 12560 |
| $T_{ver}$ | The computation time of an ECDSA signature verification based on the secp256k1 curve | 450 |
| $T_{BP}$ | Bilinear pairing | 6574 |

*A. Computation Cost*

For evaluating computational overhead, we implemented the pairing-based scheme using Pairing-Based Cryptography (PBC) and the ECC-based scheme using the Multiprecision Integer and Rational Arithmetic C/C++ Library (MIRACL). Although we used two different libraries for implementation, both were configured with a group order of 160 bits, ensuring that any discrepancies caused by the libraries were negligible. The experimental platform included an Intel i7-7500U CPU with a 2.70 GHz clock speed and 4GB RAM running on Linux Ubuntu 18.10-desktop-amd64, serving as the OBU of a vehicle. Table III lists the operations and their respective overheads.

XOR, multiplication, and arithmetic operations have been omitted. The results are shown in Table VI. All compared schemes account only for operations during mutual authentication. For our scheme, initialization and V2I authentication are excluded from the calculations. The reason for excluding these steps is that they can be optimized to occur outside of mutual authentication to reduce latency. Moreover, the protocol is primarily proposed as an efficient solution for V2V authentication, with other parts not taken into consideration. The comparison results of the computational cost for different protocols are shown in Table IV and Fig.5.

TABLE IV Comparison of Computational Cost

| Scheme | Authentication | Time(ms) |
|---|---|---|
| HDMA | $4T_{hash}+5T_{me}+2T_{BP}+2T_{dec}+T_{enc}$ | 30.311 |
| PPMA | $18T_{hash}$ | 0.036 |
| PPDAS | $3T_{hash}+2T_{sm}+T_{BP}+T_{dec}+T_{enc}$ | 15.687 |
| SPPC | $2T_{me}+8T_{hash}+2T_{sm}+3T_{dec}+3T_{enc}+T_{sig}+T_{ver}$ | 38.541 |
| EAIA | $5T_{hash}+4T_{sm}+2T_{pa}$ | 2.354 |

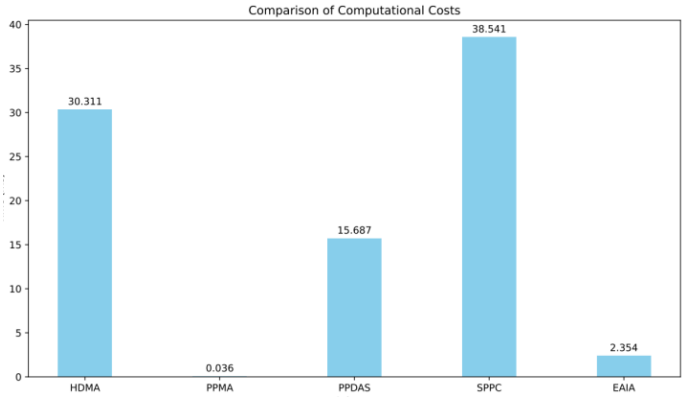

**Fig. 5** Comparison of Computational Cost

*B. Communication Cost*

The communication cost consists of transmission delay and propagation delay. According to the 3GPP specification TS 22.261 [29], in urban macro scenarios, the downlink data rate is generally 50 Mbps, and the uplink data rate is 25 Mbps. We assume a data rate of 50 Mbps between gNBs and 25 Mbps between OBUs. The transmission delay within the core network is ignored.

For propagation delay, the speed of wave propagation in wireless communication is approximately $3 \times 10^8$ m/s. Assuming a radius of 200 meters, the signal sent by a vehicle will travel 200 meters to reach another vehicle at a speed of $3 \times 10^8$ m/s. The propagation delay within the core network is also ignored. The theoretical communication cost comparison is shown in Table V, where $T_t$, $T_p$, $T_{total}$ represent transmission delay, propagation delay, and total communication time, respectively. The comparison results of the communication cost for different protocols are shown in Table V and Fig.6.

TABLE V Comparison of Communication Cost

| Scheme | Message size(bits) | Tt(μs) | Tp(μs) |
|---|---|---|---|
| HDMA | 1696 | 67.84 | 0.67 |
| PPMA | 1504 | 60.16 | 0.67 |
| PPDAS | 2272 | 90.88 | 0.67 |
| SPPC | 3216 | 128.64 | 0.67 |
| EAIA | 1312 | 52.48 | 0.67 |

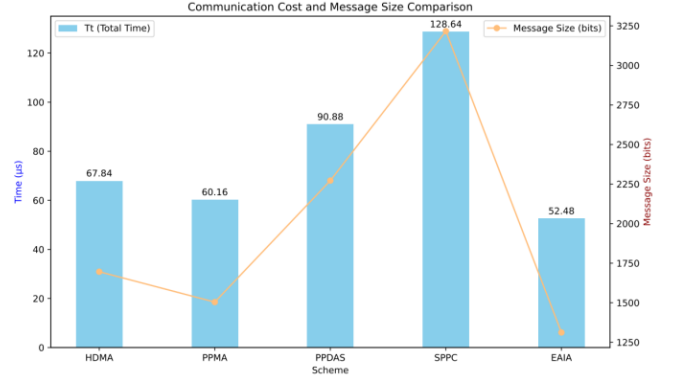

**Fig. 6** Comparison of Communication Cost

Finally, we evaluate the robustness by introducing the concept of unknown attacks. Each system is subjected to unknown attacks. When facing unknown attacks, the authentication process may be forced to stop and restart. To carefully analyze the impact of these attacks, we assume that the probability of an unknown attack occurring is uniformly distributed at each step of the authentication process. The average time for a successful handover is calculated as follows:

$$T = \frac{T_{success}+T_{failed}}{N_{success}} = \frac{\sum_{i=1}^{n}\frac{1}{n}\times t_{fail}\times p+t_{success}\times(1-p)}{1-p}$$

where $T$, $T_{success}$, $T_{failed}$ represent the average time for successful handover authentication, the total time for successful authentication, and the total time for failed authentication, respectively; $N_{success}$ denotes the number of successful authentications; p is the percentage of unknown attacks; n is the number of steps in the protocol; $t_{fail}$ indicates the amount of time spent on handover authentication before an attack occurs; and $t_{success}$ represents the time spent on successful handover authentication before an attack. The results are shown in the Fig.7.

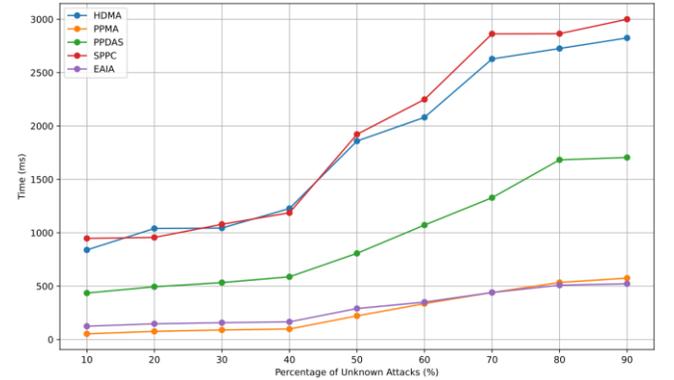

**Fig. 7**. Comparison of Authentication Time Cost

*C. Energy Consumption*

In V2V communication, energy consumption is a critical factor determining system endurance, cost-effectiveness, environmental impact, and overall sustainability. The energy consumption generated by V2V authentication includes computational energy costs and transmission energy costs. Computational energy costs primarily depend on intensive computations such as pairing calculations and scalar multiplications. For transmission energy costs, the number of messages sent and received must be considered.

In this subsection, we conduct energy analysis for computation and transmission, referring to the 133MHz SA-1110 Strong ARM microprocessor and the LA-4121 WLAN card, respectively. According to [30] and [31], the microprocessor consumes 8.8mJ for a single scalar multiplication and 47.0mJ for a bilinear pairing operation. Regarding transmission energy costs, the WLAN card consumes 0.66μJ to transmit 1 bit and 0.31μJ to receive 1 bit. Table V summarizes the energy consumption for several basic operations.

Table VI compares the energy consumption of different protocols. The results indicate that energy costs increase linearly with the number of users. Fig.8 further illustrates their energy consumption. It is evident that as the number of V2V communications increases from 5 to 100, all energy costs increase linearly. For establishing V2V group sessions, user devices consume more energy on computation than on transmission. As shown in Fig.8, the EAIA protocol demonstrates better performance in both computational and transmission energy costs. This further validates the efficiency of the proposed protocol.

## VIII. CONCLUSION

. In this paper, we designed an innovative authentication protocol based on 5G-supported V2V communication for vehicle identity authentication, specifically without the direct involvement of RSUs in the 5G wireless network. The proposed protocol has been formally verified using BAN-logic and the Scyther tool, demonstrating its security. Additionally, we analyzed its security features, showing its ability to withstand major typical malicious attacks. The proposed protocol achieves lower overhead compared to other protocols. Our experimental results indicate that this protocol outperforms its counterparts in terms of authentication overhead, providing a more efficient alternative.

TABLE VI Comparison of Energy Cost

| Notations | Description | Energy Consumption |
|---|---|---|
| $E_{me}$ | The energy cost of one exponential operation in G | 9.1mJ |
| $E_{sig}$ | The energy cost of an ECDSA (160bits) signature generation | 8.8mJ |
| $E_{ver}$ | The energy cost of an ECDSA (160bits) signature verification | 10.9mJ |
| $E_{bp}$ | The energy cost of one pairing operation | 47.0mJ |
| $E_{sm}$ | The energy cost of one scalar multiplication | 8.8mJ |
| $E_{tran}$ | The energy cost for transmitting one bit | 0.66μJ |
| $E_{rec}$ | The energy cost for receiving one bit | 0.31μJ |

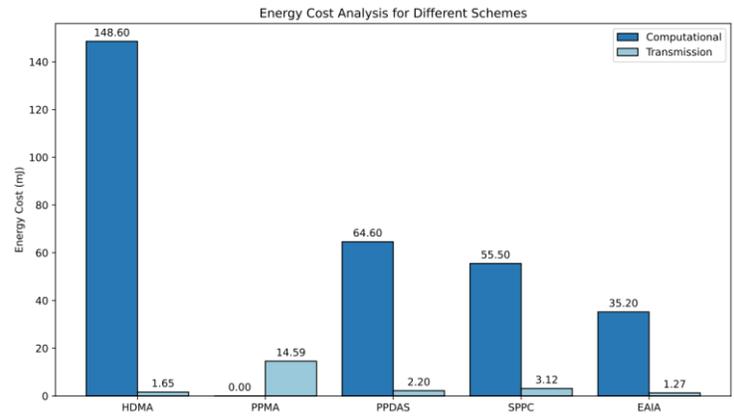

Fig. 8 Comparison of Energy Cost

TABLE VII Comparison of Energy Cost

| Scheme | Type | Operation | Energy Cost | Total Energy Cost |
|---|---|---|---|---|
| HDMA | Computational | $6T_{me}+2T_{BP}$ | 148.6 | 150.245 |
| | Transmission | 1696 | 1.645 | |
| PPMA | Computational | 0 | 0 | 14.588 |
| | Transmission | 1504 | 14.588 | |
| PPDAS | Computational | $2T_{sm}+T_{BP}$ | 64.6 | 66.804 |
| | Transmission | 2272 | 2.204 | |
| SPPC | Computational | $2T_{me}+2T_{sm}+T_{sig}+T_{ver}$ | 55.5 | 58.620 |
| | Transmission | 3216 | 3.120 | |
| EAIA | Computational | $4T_{sm}$ | 35.2 | 36.473 |
| | Transmission | 1312 | 1.273 | |